\documentclass[reprint, amsmath,amssymb, aps, nofootinbib]{revtex4-1}

\usepackage[normalem]{ulem}
\usepackage{graphicx}
\usepackage{hyperref}
\usepackage{dcolumn}
\usepackage{color}
\usepackage{bm}

\newcommand{\bq}{\begin{eqnarray}}
\newcommand{\eq}{\end{eqnarray}}
\newcommand{\ds}{\displaystyle}
\newcommand{\be}{\begin{eqnarray}}
\newcommand{\ee}{\end{eqnarray}}
\newcommand{\ba}{\begin{array}}
\newcommand{\ea}{\end{array}}
\newcommand{\pa}[1]{\left(#1\right)}
\newcommand{\paq}[1]{\left[#1\right]}
\newcommand{\pag}[1]{\left\{#1\right\}}
\newcommand{\ID}{\mathrm{ID}}

\begin{document}

\title{Deep learning waveform anomaly detector for numerical relativity catalogs}

\author{Tib\'erio Pereira}
\email{tiberio@fisica.ufrn.br}
\affiliation{Departamento de F\'isica\\Universidade Federal do Rio Grande do Norte, Natal 59078-970, RN, Brazil}

\author{Riccardo Sturani}
\email{riccardo.sturani@unesp.br}
\affiliation{Instituto de F\'\i sica Te\'orica, UNESP-Universidade Estadual Paulista \& ICTP South American Institute for Fundamental Research, Sao Paulo 01140-070, SP, Brazil}

\date{\today}

\begin{abstract}
  Numerical Relativity has been of fundamental importance for studying compact binary coalescence dynamics, waveform modelling, and eventually for gravitational waves observations. As the sensitivity of the detector network improves,
  more precise template modelling will be necessary to guarantee a more accurate estimation of astrophysical parameters.
  To help improve the accuracy of numerical relativity catalogs, we developed a deep learning model capable of detecting anomalous waveforms. We analyzed 1341
  binary black hole simulations from the SXS catalog with various mass-ratios
  and spins, considering waveform dominant and higher modes.
  In the set of waveform analyzed, we found and categorised seven types of anomalies appearing in the coalescence phases.
\end{abstract}


\maketitle

\section{Introduction}

The network of gravitational wave (GW) detectors composed by the two LIGO \cite{LIGOScientific:2014pky} and Virgo \cite{VIRGO:2014yos} observatories
has already completed three successful observation runs
\cite{LIGOScientific:2021djp}, with the detection of over 90 coalescences of compact binary systems.
To maximize the possibility of detections and their (astro)physics output,
collected data are analyzed via matched-filtering techniques
\cite{Wainstein:1962vrq,LIGOScientific:2019hgc}
by correlating them with pre-computed waveform templates, whose development
is the object of intense investigation \cite{Pratten:2020ceb,Pratten:2020fqn,Garcia-Quiros:2020qlt,Garcia-Quiros:2020qpx,Ossokine:2020kjp,Babak:2016tgq,Pan:2013rra}.
Improving the accuracy of GW templates straightforwardly enhances the quality
of astrophysical information obtained from these sources.

Waveforms can be expressed as time series resulting from the
spin-weighted spherical harmonic
decomposition of the gravitational wave strain.
While the dominant quadrupolar mode of GW templates has been enough to analyze
the vast majority of signals, in a few cases the imprints of sub-dominant
(or higher modes, HM henceforth) have been detected \cite{LIGOScientific:2020stg,LIGOScientific:2020zkf}.
With increasing sensitivity and widening the network of observatories in future
observation runs with the KAGRA observatory \cite{KAGRA:2022qtq}, it is expected that
HM will have a larger impact on the detected signals\footnote{In general
  the predicted waveform precision requirement for real events depends
  on the loudness of the signal, scaling as the inverse square of its
  signal-to-noise ratio. For current second generation detector
  \cite{Purrer:2019jcp} with signal-to-noise ratio $\sim {\rm few}\times 10$ a noise weighed ``mismatch'' of $10^{-3}$ is usually required.}.

GW waveforms generated via Numerical Relativity (NR) simulations \cite{PhysRevD.79.024003}
have been widely used for construction of semi-analytical and phenomenological
templates, and have produced accurate HM waveforms over a vast parameter space.

The present paper aims to provide a new tool to assess the data quality of the
  various waveforms presented in an NR catalog, and for the present work we
  focus on the SXS catalog \cite{Boyle:2019kee}.
  The numerical differentiation and time integration methods can generate a systematic accumulation of numerical residue \cite{Boyle:2006ne, Rinne:2007ui, Szilgyi:2014} -- even more for extreme regimes such as binary black holes simulations -- leading to defects in the morphology of the waveforms. Furthermore, there are cases where a catalog adds simulations with improvements in resolution (or with better numerical methods) and does not remove old simulations -- because they still prove useful for theoretical studies.
Hence such waveforms are the natural candidates to be identified by our tool
as the ones to display inconsistencies within the catalog.

In this work, we developed the deep learning model \emph{Waveform AnomaLy DetectOr} (WALDO), capable of signaling possible anomalous waveforms in a NR catalog \cite{zenodo, github, ascl}.
In our searches within binary black hole (BBH) simulations, we categorized seven different types of
anomalies during the stages of coalescence. Identifying and excluding such
waveforms is critical to the quality of research in GW analysis and surrogate
modeling \cite{Varma:2019csw}.

Applications of deep learning models to gravitational wave data is not new
\cite{Easter:2018pqy,Gabbard2018,Varma:2018aht,Shen2019,Rebei2019,Setyawati:2019xzw,Haegel:2019uop,Cuoco:2020ogp,Green2020,Ormiston2020,2111.03295,Schmidt2021,Gabbard2021,2203.08434,Yan2022}, 
but to the best of our knowledge this is the first
work using deep learning to check the consistency of numerical simulations.

The paper is structured as follows. Section \ref{sec:defs} is intended
to help the reader providing a reference to our notations, Section \ref{sec:data}
describes the dataset we used for our analysis, and Section \ref{sec:waldo}
describes our machine learning-based process to identify anomalous waveforms,
whose results are presented in Section \ref{sec:results}.
Finally we summarize our conclusions in Section \ref{sec:concl}.

\section{Definitions}
\label{sec:defs}
We adopt geometric units $G=c=1$, and we denote by $u$ the dimension-less
time obtained by dividing physical time by the total mass $M$ of the binary
system, whose zero is set by the epoch of the peak of the dominant mode amplitude.
The BBH mass-ratio $q$ is taken to be larger than 1, the dimensionless spins
$\vec\chi_i \equiv \vec S_i/M^2$, $i = \{1, 2\}$, being $\vec S_i$ the standard
spin and the orbital eccentricity is denoted by $e$.
From the GW \emph{strain}, i.e. the GW polarization complex combination
$h_+-ih_\times$, we extract
(and rescale as usual by distance $r$ and mass $M$)
spherical harmonic modes $h_{lm}(t)$, indexed by integers
$l\geq 2$ and $|m|\leq l$, resulting from the decomposition on the spin-weighted
spherical harmonics base $_{-2}Y_{lm}$ of the strain,
\be
h_{lm}(t) = \int d\Omega\, \pa{h_+-ih_\times}(\theta,\phi)\,_{-2}Y_{lm}^*(\theta,\, \phi)\,,
\ee
where $\Omega$ is the solid angle parameterized by $\theta$ and $\phi$,
which are respectively the angle between the radiation direction and the normal
to the orbital plane, and a phase corresponding to a rotation in the orbital
plane.

\section{The dataset}
\label{sec:data}
\begin{figure}[h]\centering
 \hspace*{-0.4cm}
 \includegraphics[width=8cm]{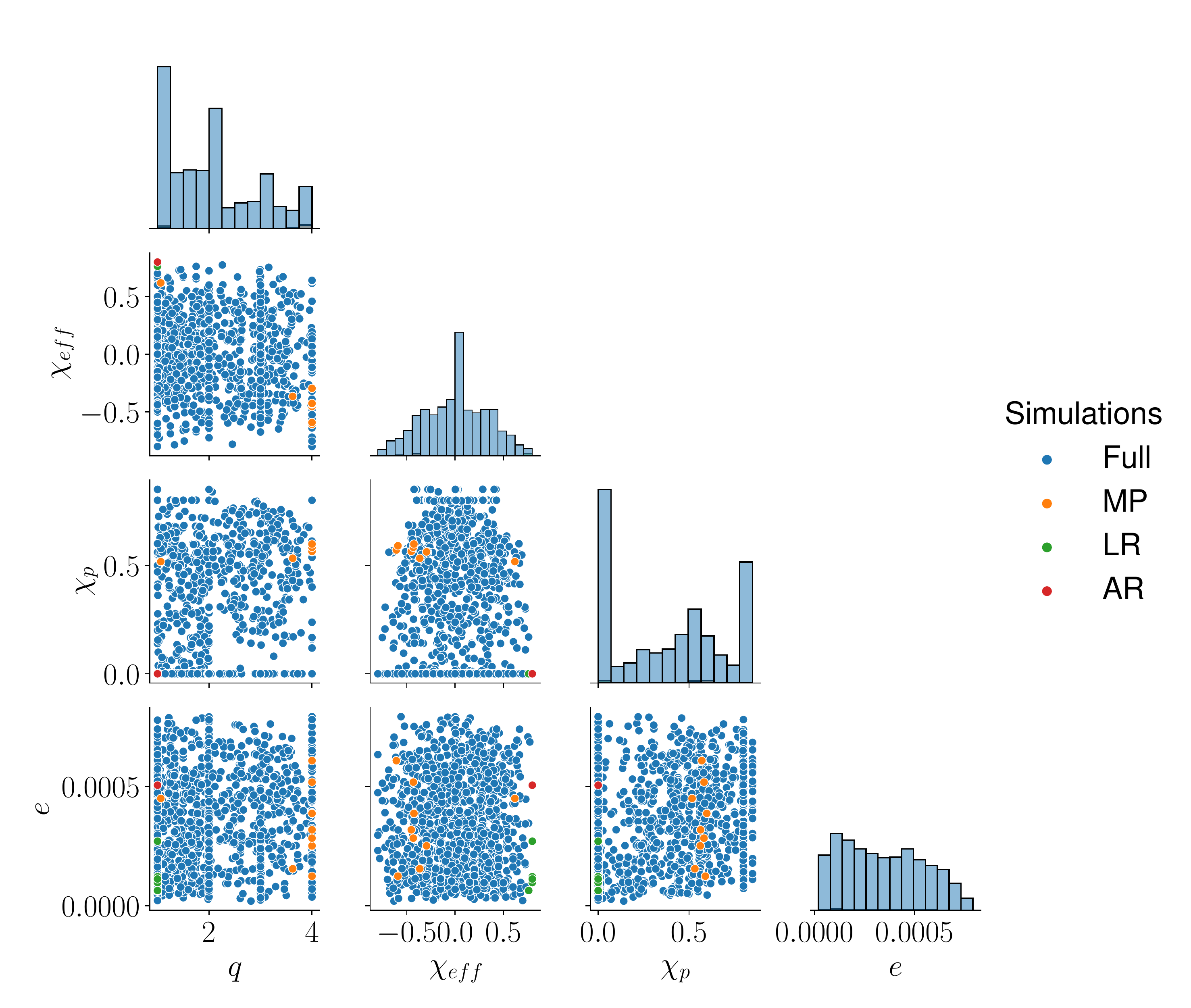}
  \caption{Representation of the parameter space of the 1341 simulations: corner
    plot of the distribution of eccentricity $e$, mass-ratio $q$, spin-aligned
    parameter $\chi_{eff}$ and spin-precession $\chi_p$, defined in Sec. \ref{sec:data}. We highlighted the waveforms presenting the anomalies displayed
      in the list on the right of the figure, whose acronyms are defined in Tab. \ref{tab:anom}.}
 \label{fig:paramsf}
\end{figure}

We create a dataset using 1341 BBH simulations from the \emph{Simulating eXtreme Spacetimes} (SXS) catalog \cite{PhysRevD.79.024003}, whose parameters
are in the region
$q = [1.0:4.0]$, $|\vec\chi_i| = [-0.9:0.9]$, and $e \simeq [0.0:8.0\times10^{-4}]$ \cite{Mrou2013,Boyle:2019kee}.\footnote{We restricted our analysis to the
    1341 simulations with mass ratio $q\leq 4$ because beyond this value
    simulations become sparser, and this can jeopardize the learning
    process of the neural network model training.}
All simulation names are listed in the WALDO's repository \cite{zenodo} (\texttt{simulations\_ID.txt} file).
Considering the modes $(l\leq 4,\, l-1\leq m\leq l)$\footnote{To make it explict,
  we use the $(2,2),\,(2,1),\,(3,3),\,(3,2),\,(4,4),\,(4,3)$, \emph{i.e.}, 6 modes for each simulation in the catalog.}, in total our dataset is composed by $N_d=6\times 1341=8046$ waveforms. 
Figure \ref{fig:paramsf} shows the parameter space distribution of eccentricity, spin-aligned parameter $\chi_{eff}$ \cite{Ajith:2009bn}
spin-precession parameter $\chi_p$ \cite{Schmidt:2014iyl} and the
mass-ratio $q>1$, highlighting the simulations showing anomalies (discussed in the Section \ref{sec:results}).
The spin parameters $\chi_{eff,p}$ are approximately constant
  even in the case of precession, so they are particularly useful in
  characterizing the waveform parameter space. Their explicit expressions are
  \be
  \ba{rcl}
  \ds\chi_{eff}&\equiv &\ds\frac{S_{1\parallel}/m_1+S_{2\parallel}/m_2}{m_1+m_2}\,,\\
  \ds\chi_p&\equiv &\ds \frac{{\rm max}\pa{A_1 S_{1\perp}+A_2 S_{2\perp}}}{A_2m_2^2}\,,
  \ea
  \ee
  where $S_{i\parallel,\perp}$ denote the spin component of the $i$-th binary constituent respectively parallel and perpendicular to the Newtonian orbital angular
  momentum, with $A_i\equiv 2+3m_i/(2m_{j\neq i})$.

To facilitate a unified treatment of all waveforms, we cut the waveforms to the highest
initial time value of the entire dataset, \emph{i.e.}, the inspiral starts at $u=-1700$, with corresponding instantaneous
frequency for the dominant mode $f_{min} \simeq 14-19 Hz\pa{M/100M_\odot}^{-1}$,
depending on the mass ratio. This conditioning is necessary to guarantee the same resolution of the waveforms -- with the same time intervals -- during the neural network (NN) training \footnote{For all numerical waveforms we used the best resolution available in the repository.}. 

Also, we find it convenient to re-sample all modes via the time-reparametrization
\be
\begin{array}{rcl}
\label{eq.time}
u'(u) &=&\ds c_1\,r(u) + c_2\,,\\
r(u) &=&\ds \tanh\paq{a\pa{u - u_0 + b}}\,,\\ 
c_1 &=&\ds \pa{u_F-u_0}/\paq{r(u_F)-r(u_0)}\,,\\
c_2 &=&\ds u_0-c_1\,r(u_0)\,,
\end{array}
\ee
where $u_0$ and $u_F$ are the initial and final value of the
dimension-less time $u = [-1700:100]$; $a=5.0\times10^{-4}$ and $b=5.0\times 10^2$ are constants, see \emph{e.g.} \cite{Schmidt:2020yuu} for analog reparametrization.
The rationale for this parameterization is to make smoother
the transition from the wider spacing during the inspiral to a smaller one
in the merger-ringdown phase,
while keeping the number of samples
equal to $N_s=2048$ for all waveforms, without degrading the sampling rate
in the merger-ringdown phase.
This reparameterization does not bring morphological issues or degenerate the accuracy of the waveforms, 
but it helps the NN to capture more precise features after the inspiral.

For deep learning \emph{feature engineering} -- the pre-processing procedures for improving NN computations --  we normalize the entire dataset with the highest peak value among all waveform amplitudes, 
\be
    h_{k,lm} &\rightarrow& h_{k,lm}/\max\pa{A_k} \,,
\ee
where the index $(k)$ denotes the simulation number, $k=[1:1341]$, and
$A_k$ is the dominant mode maximum amplitude value of the $k$-th simulation.
We define the dataset as the numerical three-dimensional array,  
\be
    {\bf X} &\equiv& \pa{\mathit{Re}\left\{h_{k,lm}\right\}, \mathit{Im}\left\{h_{k,lm}\right\}}\label{input}\,,
\ee
whose dimensionality is $(N_d,\, N_s, 2)$.

\section{WALDO} 
\label{sec:waldo}
The Waveform AnomaLy DetectOr (WALDO) holds a U-Net architecture \cite{1505.04597}, where the waveform input $X_{k,lm}$ is reproduced as the output $\bar X_{k,lm}$. During the training, the model learns all possible waveform features related to the parameter space. Evaluating its performance after training
with the \emph{mismatch} $\mathcal{M}$
between $h_{k,lm}$ and its prediction $\bar h_{k,lm}$, the measurement
of high mismatch values can flag the presence of waveforms whose morphology
do not match the dataset one, \emph{i.e.}, we can find \emph{anomalous waveforms}.
The mismatch is defined as $\mathcal{M}\equiv 1-\mathcal{O}$, where
${\cal O}$ denotes the \emph{match}
between two time series $h_{1,2}$, defined in terms of the scalar product
\be
\langle h_1|h_2\rangle = 4\int_0^{\infty} \frac{\tilde h_1(f)\tilde h_2^*(f)}{S_n(f)} df\,,\label{eq:snr}
\ee

where $f$ is variable conjugate to time under Fourier transform $h\to \tilde h$, and $S_n(f)$ is the noise spectral density -- we set $S_n(f) = 1$ for noise-free waveforms.
The match being defined by maximization over initial phase and time
of the scalar product of normalized waveforms 
\be
\mathcal{O}\equiv \stackrel{Max}{t,\phi_0}\frac{\langle h_1| h_2\rangle}
{\pa{\langle h_1|h_1\rangle\langle h_2|h_2\rangle}^{1/2}}\,.
\ee

\begin{figure}[h]\centering
\hspace*{-0.5cm}
 \includegraphics[width=8.0cm]{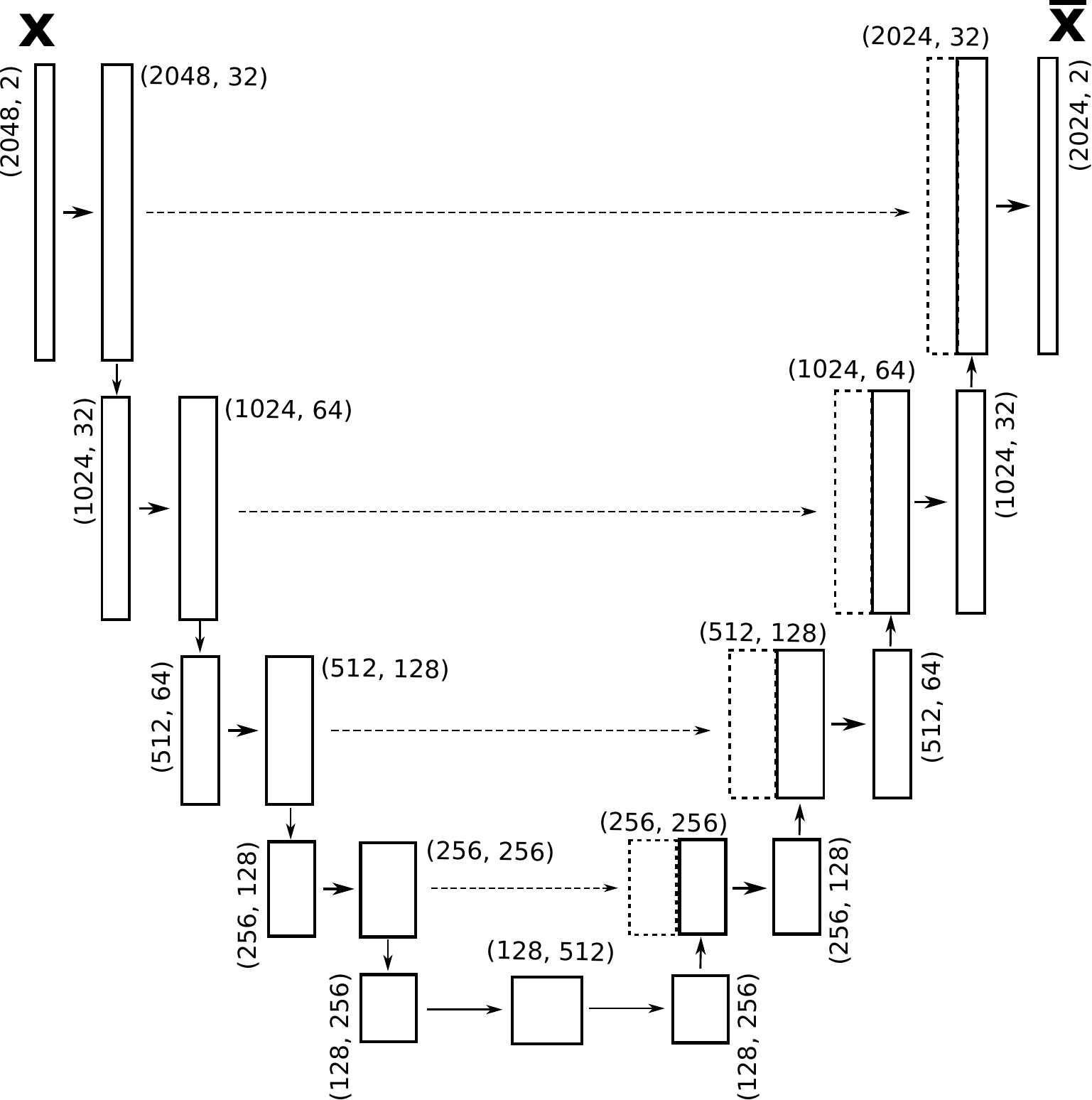}
 \caption{Illustration of the WALDO's U-Net architecture.}
 \label{unet}
\end{figure}

The WALDO's architecture is illustrated in Fig. \ref{unet}, where the input for one waveform, addressed by Eq. \ref{input}, undergoes a succession of convolution and max-pooling operations given by the solid right and down arrows, respectively. Those operations in the encoder part (the left-hand side blocks) use a one-dimensional kernel/window to sweep the time series and make operations for output. We adopt a $\emph{kernel-size}=3$ for convolutional layers and the respective number of kernels $C$ per encoder layer as $C = [32,\, 64,\, 128,\, 256,\, 512]$. For the max-pooling layers, we use a $\emph{pool-size}=2$ to halve the time series size (by taking the maximum value between two points) as the number of convolutional kernels increases to improve the feature extraction.

The model's decoder part (the right-hand side blocks) doubles the time series size with up-sampling layers, indicated by up arrows. It concatenates their outputs with the encoder layer with the corresponding dimensions, shown by dashed right arrows and blocks. These operations follow a convolutional layer which reduces the number of kernels. In short, the decoder blocks restore the input {\bf $X$} original dimension meanwhile transfer the encoder features to improve {\bf $\bar X$} resolution. We compose the whole architecture with ReLU activation functions except for the hyperbolic tangent in the output layer.

An unsupervised model for the anomaly detection task usually uses an auto-encoder architecture, a structure similar to the scheme of Fig. \ref{unet} but without the connections represented by the dashed arrows. In the U-Net model, these connections help the convolutional layers of the decoder to reproduce the kernels that best extract the features in the encoder blocks. Therefore, the U-Net model can perform better in waveform reproduction than a conventional autoencoder model for a reduced dataset. This choice with the Adagrad optimizer allows a smooth minimization of the loss function and high resolution waveform reproduction.

\subsection{Training and validation}
\label{ssec:train_val}
\begin{figure}[h]\centering
\hspace*{-0.6cm}
 \includegraphics[width=9.0cm]{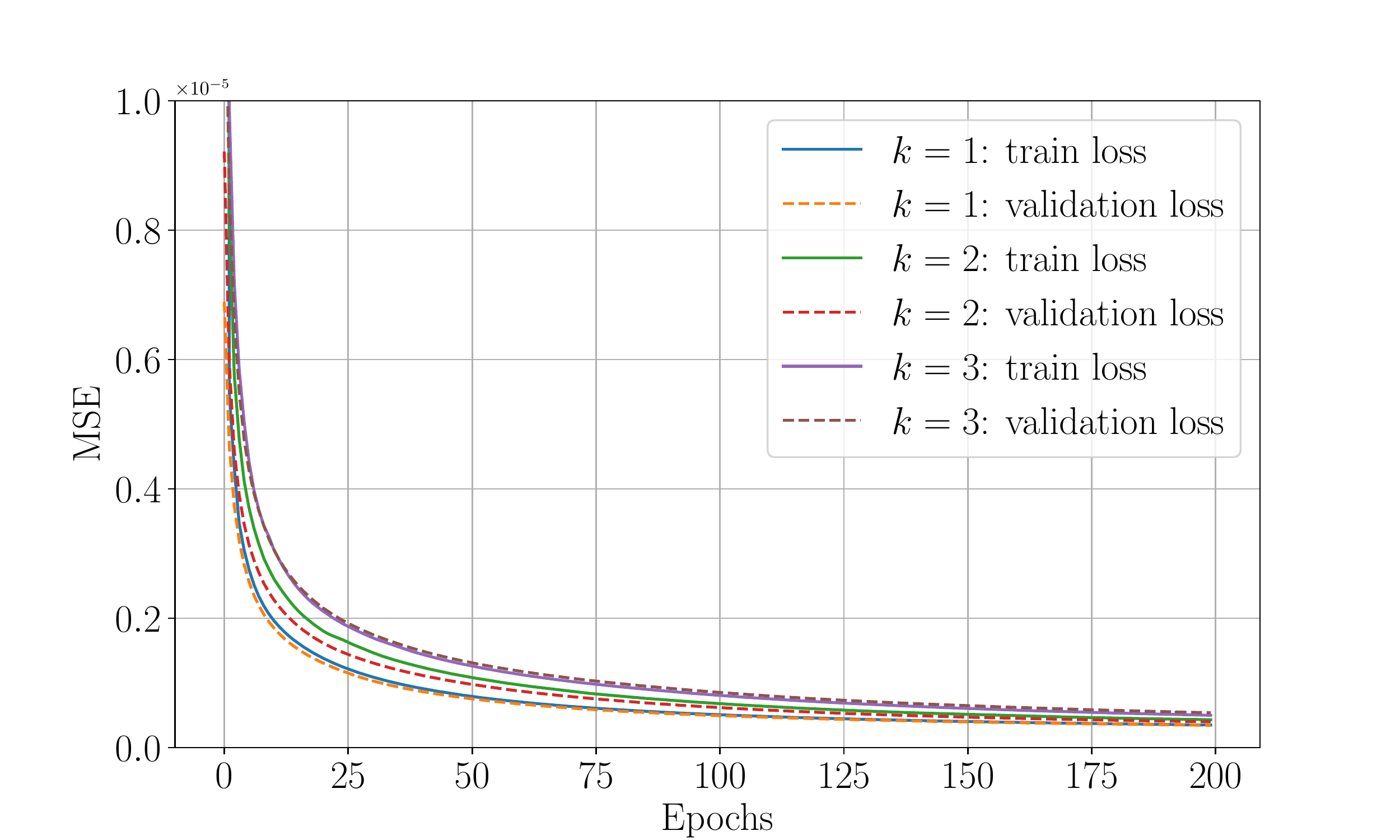}
 \caption{The mean squared error (MSE) evaluation over 200 epochs in 3-fold validation performance. Solid lines represent the training loss, while dashed ones show the validation loss.}
 \label{kfold}
\end{figure}

To examine the training performance, we split the dataset into 90\% for training and validation data and 10\% for testing data. Since 8046 waveforms form a small dataset, we use the \emph{K-fold validation} method for $K = 3$ and $\emph{batch-size} = 1$. We optimize the model parameters using the mean squared error (MSE) loss function and Adagrad optimizer \cite{DuchiHS11}. During the validation, the NN can be trained through 200 epochs without over-fitting since we applied an early-topping mechanism to prevent the training for the difference of train and validation losses greater than $1.0\times 10^{-8}$. We retrain the model using 90\% of the dataset and obtain the MSE average of $2.8\times 10^{-7}$ over the testing data. Figure \ref{kfold} shows the training and validation performance with, respectively, solid and dashed lines per each fold-validation.

\section{Results}
\label{sec:results}
\begin{figure}[h]\centering
\hspace*{-0.5cm}
 \includegraphics[width=8.8cm]{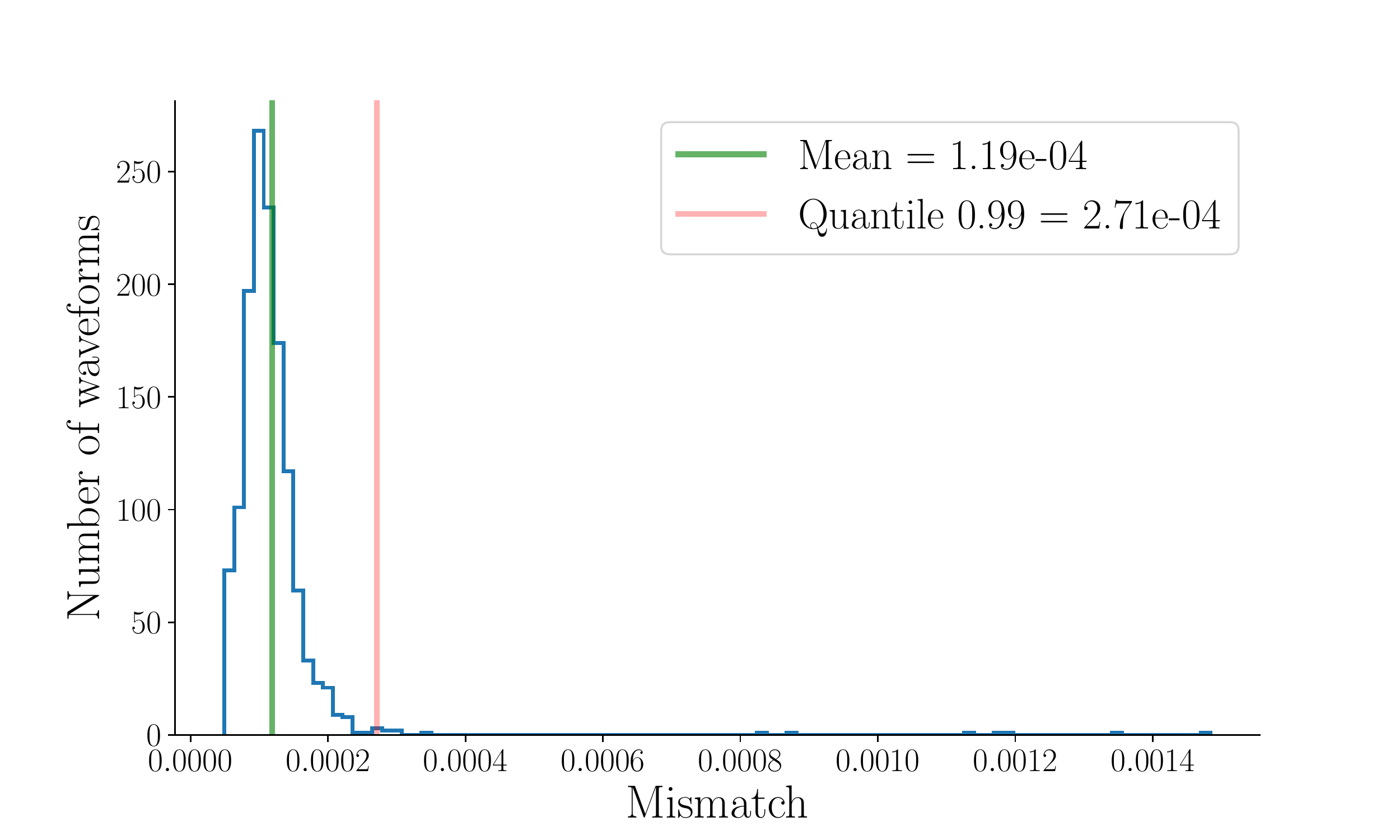}
    \caption{Mismatch histogram of 1341 waveforms $h_{32}^{(k)}$ and the predictions $\bar h_{32}^{(k)}$. The vertical green line shows the average value $1.26\times 10^{-5}$, and the pink one marks the $quantile=0.99$ with 14 simulations on its right side.}
 \label{hist}
\end{figure}

After the training, WALDO evaluates the mismatch and packs the values with $(l,\,m)$ mode labels, together with the identification simulation number (ID) -- that comes from the SXS simulation names as SXS:BBH:ID -- the parameter space $\pa{q,\, \vec\chi_1,\, \vec\chi_2,\, e}^{(k)}$, the waveforms $h_{k,lm}$ and their predictions $\bar h_{k,lm}$. Creating a histogram for each mode $(l,\,m)$, we isolated 1\% of the highest mismatch waveforms to verify any possible morphological discrepancy in the predictions. Figure \ref{hist} shows the $(3,\, 2)$ waveform mismatch distribution of average $1.26\times 10^{-5}$ represented by the vertical green line; the $quantile = 0.99$, marked by the vertical pink line, separates 14 simulations that call for examination. 
The lowest mismatch value is on the order of $5.0\times 10^{-6}$ due to the low number of simulations $N_d=1341$ and varied waveform morphology.
We reinforce that the NN training is usually done with hundreds of thousands of data; however, our small dataset does not interfere with the quality of waveform reproduction once our NN model achieved a mismatch average of $1.25\times 10^{-5}$, while the SXS threshold accuracy is $1.0\times 10^{-5}$ times above the amplitude peak \cite{sxs_site}.

\begin{figure}[h]\centering
\hspace*{-0.2cm}
 \includegraphics[width=8.5cm]{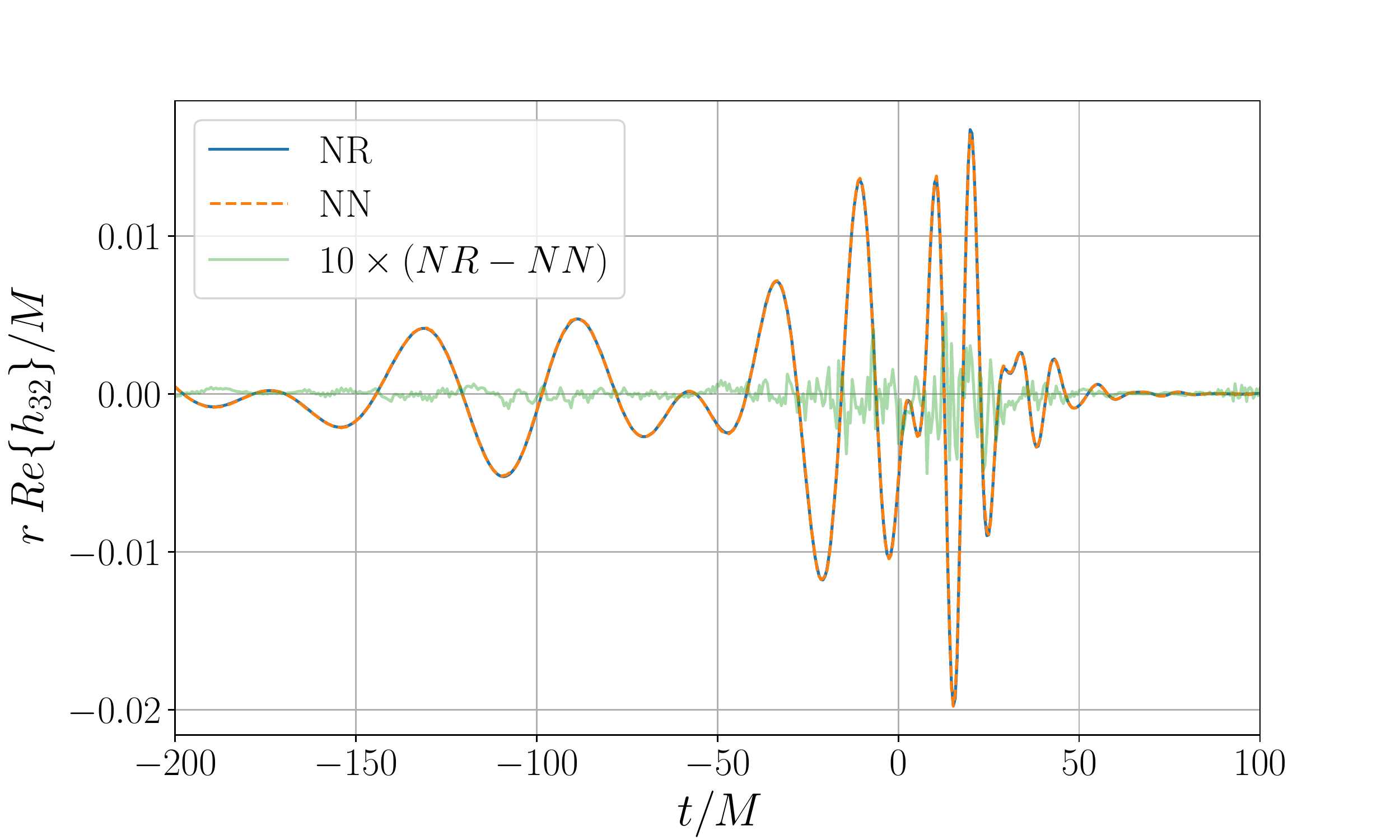}
    \caption{NR vs NN comparison of the $(3,\, 2)$-waveforms from the simulation ID = 1018.
      $Mismatch = 3.0\times 10^{-4}$; $q = 1.2$, $\vec\chi_1 = (-0.59, 0.15, -0.39)$, $\vec\chi_2 = (-0.73, 0.19, 0.22)$, and $e = 4.2\times 10^{-4}$.}
 \label{noisy_wf}
\end{figure}

In this case, we found high mismatches due to noise accumulation in the
predicted waveforms, even if qualitatively they follow the NR morphological
patterns, as shown in Fig. \ref{noisy_wf} -- where the blue line is $h_{32}$, the dashed orange line is $\bar h_{32}$, and the green one is $Re\pag{h_{32}-\bar h_{32}}$ amplified 10 times. On the other hand, we also found morphological
discrepancies between NN predicted and NR waveform modes,
confined in specific sectors of the coalescence, causing high mismatches.

\begin{figure}[h]\centering
\hspace*{-0.2cm}
 \includegraphics[width=8.5cm]{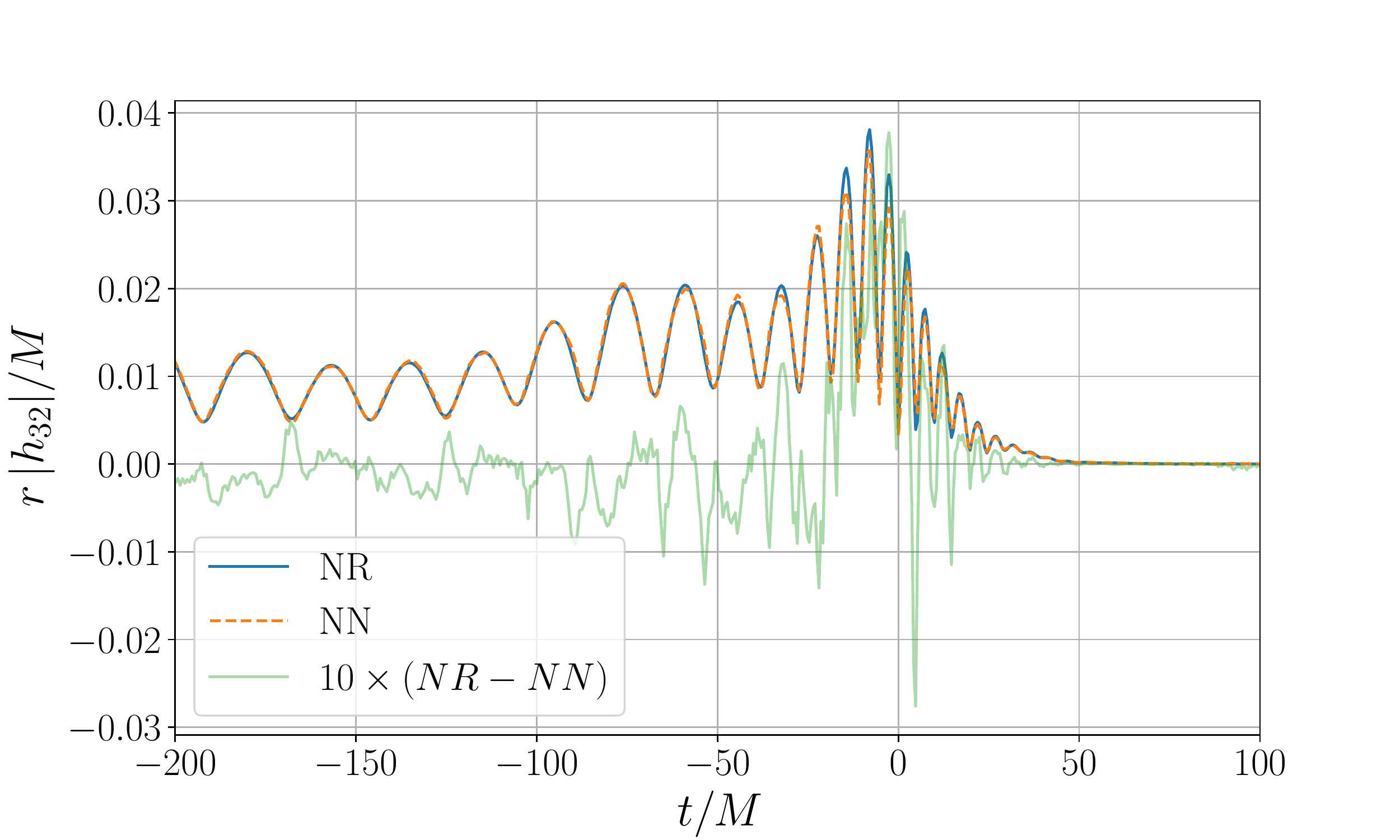}
    \caption{NR vs NN comparison of the $(3,\, 2)$-waveforms from the simulation ID = 1995, displaying a MP anomaly.
      $Mismatch = 1.48\times 10^{-3}$; $q = 4.0$, $\vec\chi_1 = (0.51, -0.29, -0.54)$, $\vec\chi_2 = (0.07, 0.07, -0.79)$, and $e = 1.2\times 10^{-4}$.}
 \label{MP_plot}
\end{figure}

An irregularity fairly common for the higher modes shows that the mode amplitude around the merger peak has a greater magnitude than expected -- what we call the \emph{merger-peak} (MP) anomaly, as seen in Fig. \ref{MP_plot}.

For even $m$ modes, we found MP anomalies in $\ID_{(3,2)} = [1863,\, 1991-1993,\, 1995]$ and $\ID_{(4,4)} = [1982,\, 1983,\, 1991,\, 1992,\, 1995]$. In the search for odd $m$ modes, we restrict the mass ratio to $q > 1.0$, where we found $\ID_{(2,1)} = [1982,\, 1993]$, $\ID_{(3,3)} = [1982,\, 1983,\, 1989, 1991-1993,\, 1995]$, and $\ID_{(4,3)} = [0601,\, 1982,\, 1983,\, 1989,\, 1991-1993,\, 1995]$.

\begin{figure}[h]\centering
\hspace*{-0.2cm}
 \includegraphics[width=8.5cm]{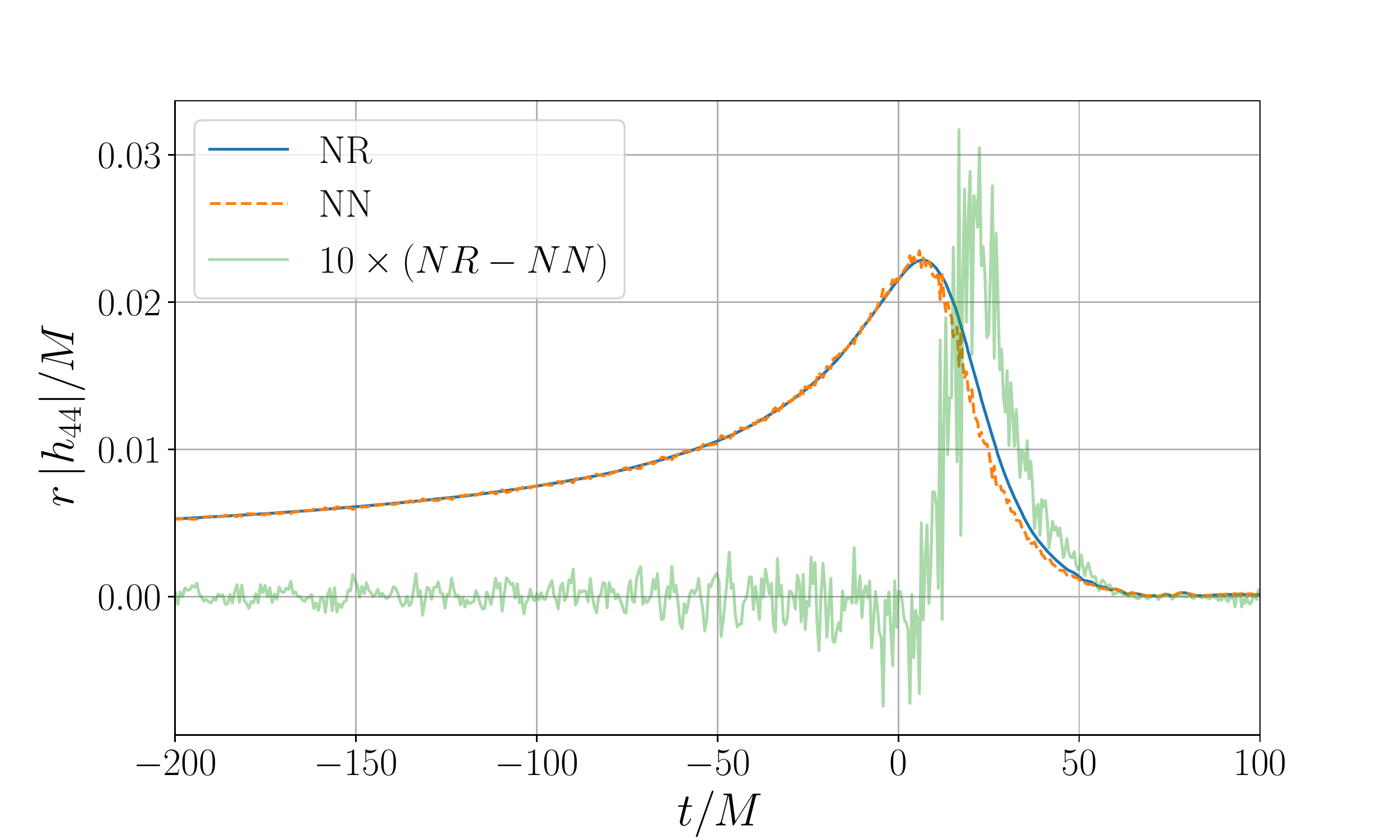}
 \caption{NR vs. NN comparison of the $(4,\,4)$-waveforms from the simulation ID = 0328 displaying a LR anomaly.
   $Mismatch = 1.44\times 10^{-3}$; $q = 1.0$, $\vec\chi_1 = (0.0, 0.0, 0.8)$, $\vec\chi_2 = (0.0, 0.0, 0.8)$, and $e = 1.1\times 10^{-4}$.}
 \label{LR_plot}
\end{figure}

\begin{figure}[h]\centering
\hspace*{-0.2cm}
 \includegraphics[width=8.5cm]{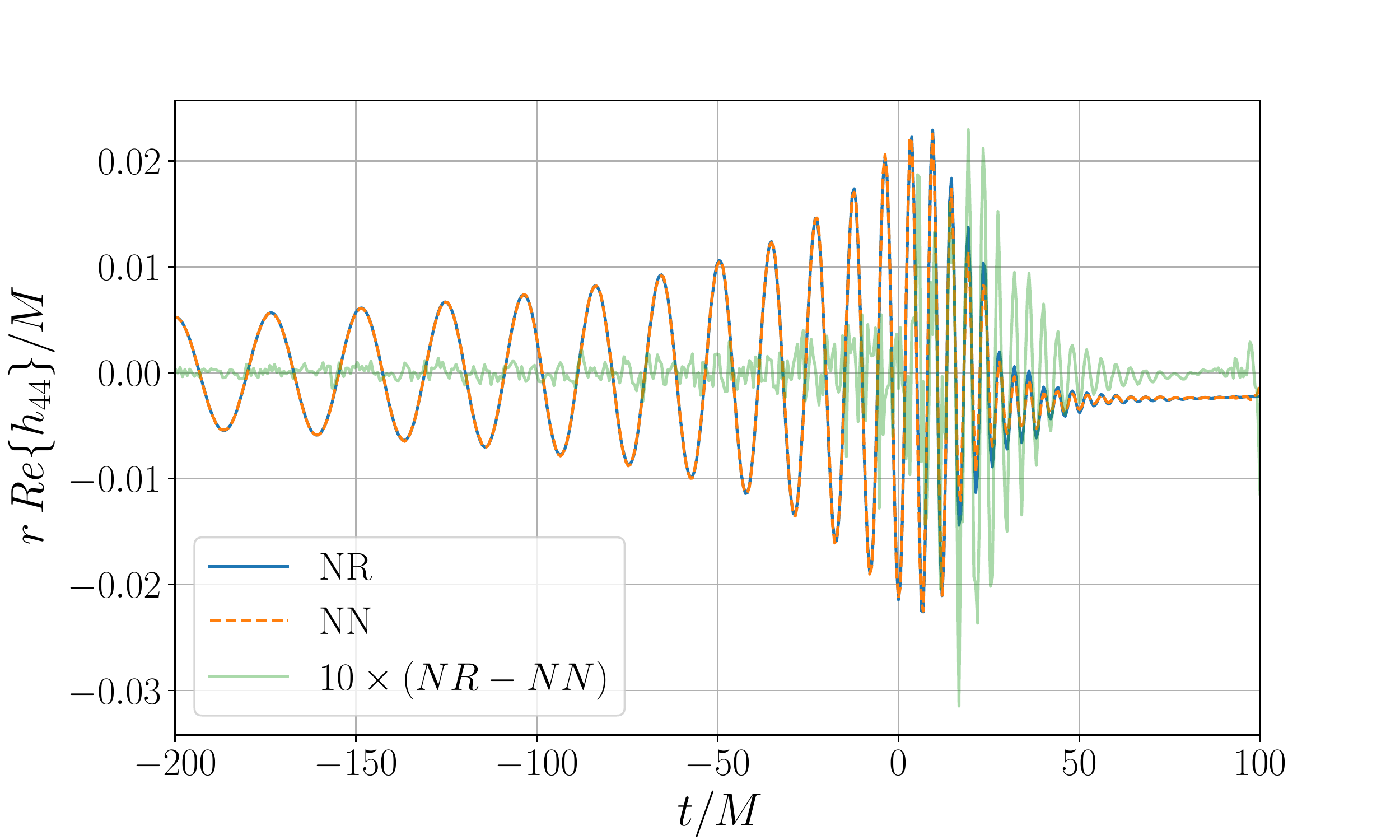}
    \caption{NR vs. NN comparison of the $(4,\, 4)$-waveforms from the simulation ID = 0155 exemplifying a AR anomaly.  $Mismatch = 1.21\times 10^{-3}$; $q = 1.0$, $\vec\chi_1 = (0.0, 0.0, 0.8)$, $\vec\chi_2 = (0.0, 0.0, 0.8)$, and $e = 5.1\times 10^{-4}$.}
 \label{AR_plot}
\end{figure}

In some waveform $(4,\,4)$ modes the ringdown decay begins a little later
in NR simulations than in our NN predictions. Figure \ref{LR_plot} shows an
example of the \emph{lazy-ringdown} (LR) anomaly, also found in $\ID_{(4,\,4)} = [0230,\, 1477,\, 1481,\, 2104]$. In the simulation $\ID_{(4,4)} = 0155$, on the other hand, the ringdown amplitude does not exhibit appropriate asymptotic behavior. The \emph{asymptotic-ringdown} (AR) anomaly is shown in Fig. \ref{AR_plot}.

Figure ~\ref{fig:paramsf} shows the parameter space corner plot of all simulations
  as blue dots, highlighting the simulations with Merger-Peak (MP, orange),
  Lazy-Ringdown (LR, green), and Asymptotic-Ringdown (AR, red). We can notice
  that we only have AR and LR anomalies in the very boundary of the parameter
  space. To check whether these anomalies are due to the NN learning bias, we
  constrain the search in the dataset for further analysis.

\subsection{Constrained dataset search}
\label{ssec:const}
The homogeneity of the parameter space distribution is important to avoid
WALDO prediction bias. For instance, a dataset containing 500 simulations of spin-aligned BBH and 20 precessing binaries can lead to high mismatch values for waveforms whose features indicate precession. 

\begin{figure}[h]\centering
\hspace*{-0.2cm}
 \includegraphics[width=8.5cm]{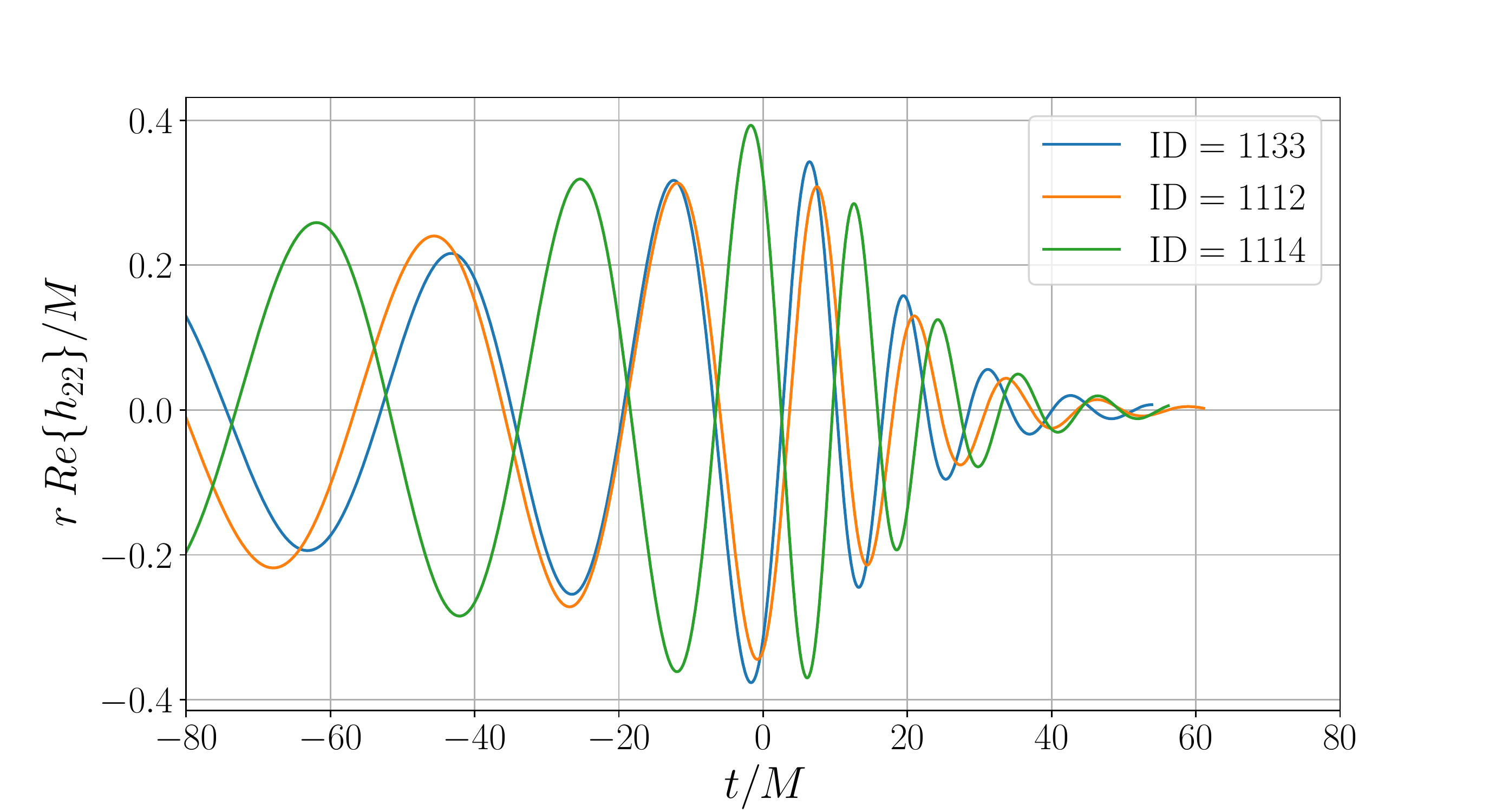}
 \caption{Short-ringdown anomaly in the simulations ID = [1112, 1114, 1133].}
 \label{SR_plot}
\end{figure}

Thereby, we focus our search for anomalies on $|\vec\chi_{1,2}| \leq 0.4$ -- but keeping the same training set -- where we have a higher simulation density. We choose $quantile=0.90$
to isolate 15 waveforms. This constraint leads us to find more AR anomalies in $\ID_{(3,\,3)} = [0116,\, 0115,\, 0119,\, 0129]$ and $\ID_{(4,\,4)} = [0070,\, 0115,\, 0124,\, 0135,\, 0150]$. In addition, we found waveforms with similar decay as in Fig. \ref{AR_plot}, but with non-oscillatory patterns. In this case, those ringdowns were affected by the time interpolation of Eq. \ref{eq.time} because
their final time is smaller than $u = 70$, giving rise to what we dubbed
\emph{short-ringdown} (SR) anomaly, found in $\ID_{(l,m)} = [1112,\, 1114,\, 1133]$, for all $(l,m)$. Fig.~\ref{SR_plot} shows the $h_{22}$ modes from these
simulations.

\begin{figure}[h]\centering
\hspace*{-0.2cm}
 \includegraphics[width=8.5cm]{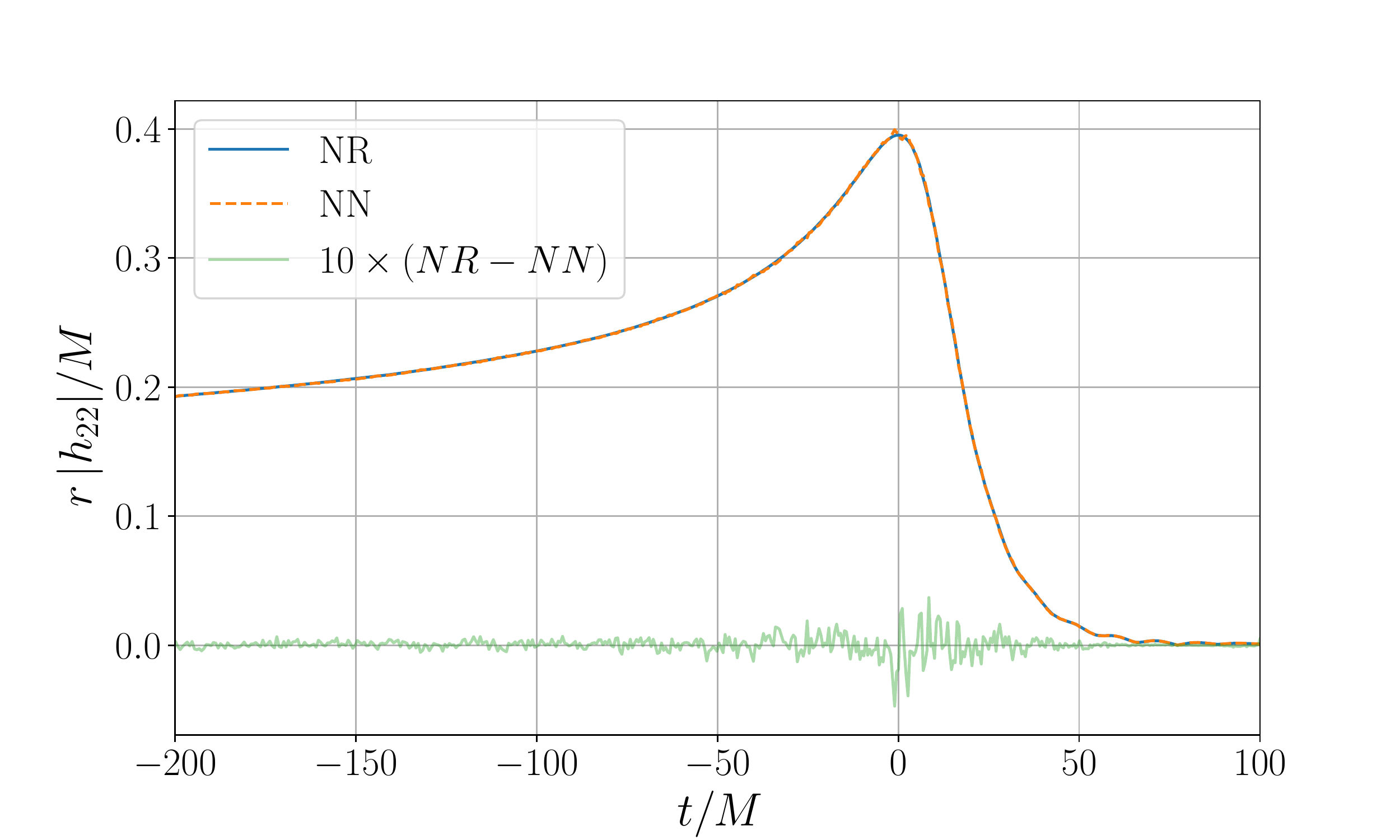}
 \caption{NR vs. NN comparison of the $(2,\, 2)$-waveforms from the simulation ID = 0072, showing a RR anomaly. $Mismatch = 1.33\times 10^{-5}$ ; $q = 1.0$, $\vec\chi_1 = (0.0, 0.0, 0.0)$, $\vec\chi_2 = (0.0, 0.0, 0.0)$, and $e = 1.5\times 10^{-4}$.}
 \label{RR_plot}
\end{figure}

The ringdown amplitude of the dominant mode is expected to show a smooth,
quasi-exponential decay, however, in the simulations $\ID_{(2,2)} = [0066,\, 0067,\, 0070,\, 0072,\, 0126,\, 0136]$ appear small ripples up to $u = 40$ as in
Fig.~\ref{RR_plot}, which we call the \emph{rippled-ringdown} (RR) anomaly.

Note that these small ripples are present in several of the original NR simulations, and they
are reproduced by the NN predictions, however with high enough mismatch to
be uncovered.

\begin{figure}[h]\centering
  \hspace*{-0.4cm}
  \includegraphics[width=8cm]{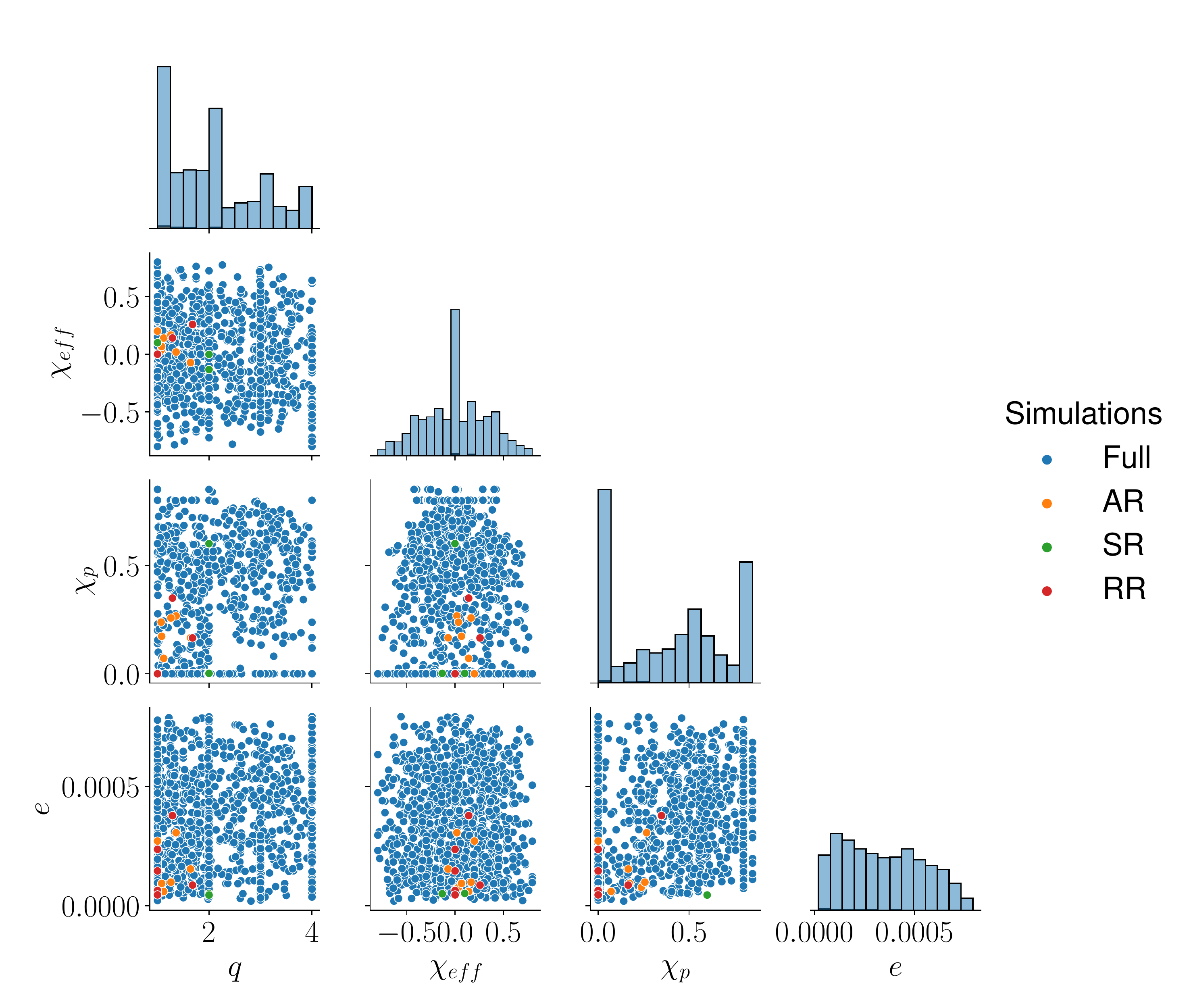}
  \caption{Analog to Fig.~\ref{fig:paramsf}, but highlighting only
      the anomalies of the constrained search of the waveforms, see Tab.~\ref{tab:anom} for
      acronyms.}
  \label{fig:paramsc}
\end{figure}

The simulations analyzed in the constrained search are not on the parameter space boundary we chose (apart from the $q=1$ physical boundary), see Fig. \ref{fig:paramsc}. By focusing on our previously defined constrained set of parameters, the AR anomalies are the only ones which are carried over from previous analysis, indicating that their presence is not due to NN learning bias due to parameter space boundaries.

\subsection{Radiation field search} 

In NR the strain modes are usually computed by the integration of the radiation field $\Psi_4$ \cite{Reisswig2011}, or independently, via perturbative formulation of the metric -- the Regge-Wheeler-Zerilli formalism used in SXS simulations \cite{Ruiz2007,Nagar2006}. We investigate the $\psi_{lm} = \ddot h_{lm}$ time series within the whole dataset to ensure the quality of both strain and $\Psi_4$ modes.  

We retrained the NN, using the same architecture setup as before and training time, and evaluate the mismatch between NR $\psi_{k,lm}$ waveforms and their reproductions $\bar\psi_{k,lm}$.

From previous analyses, we found exclusively MP anomalies in $\psi_{lm}$ modes in
$\ID_{(2,2)} = [1993,\, 1995]$, $\ID_{(3,2)} = [1863,\, 1878,\, 1983,\, 1991-1993,\, 1995]$, $\ID_{(3,3)} = [1982,\, 1989,\, 1991-1993]$, $\ID_{(4,3)} = [1103,\, 1982,\, 1983,\, 1989,\, 1991-1993]$, and $\ID_{(4,4)} = [0024, 0126, 1982, 1989, 1993, 1995]$. 
Some of these simulations listed are the same as those in the previous $h_{lm}$ waveform analysis.

\begin{figure}[h]\centering
\hspace*{-0.2cm}
 \includegraphics[width=8.5cm]{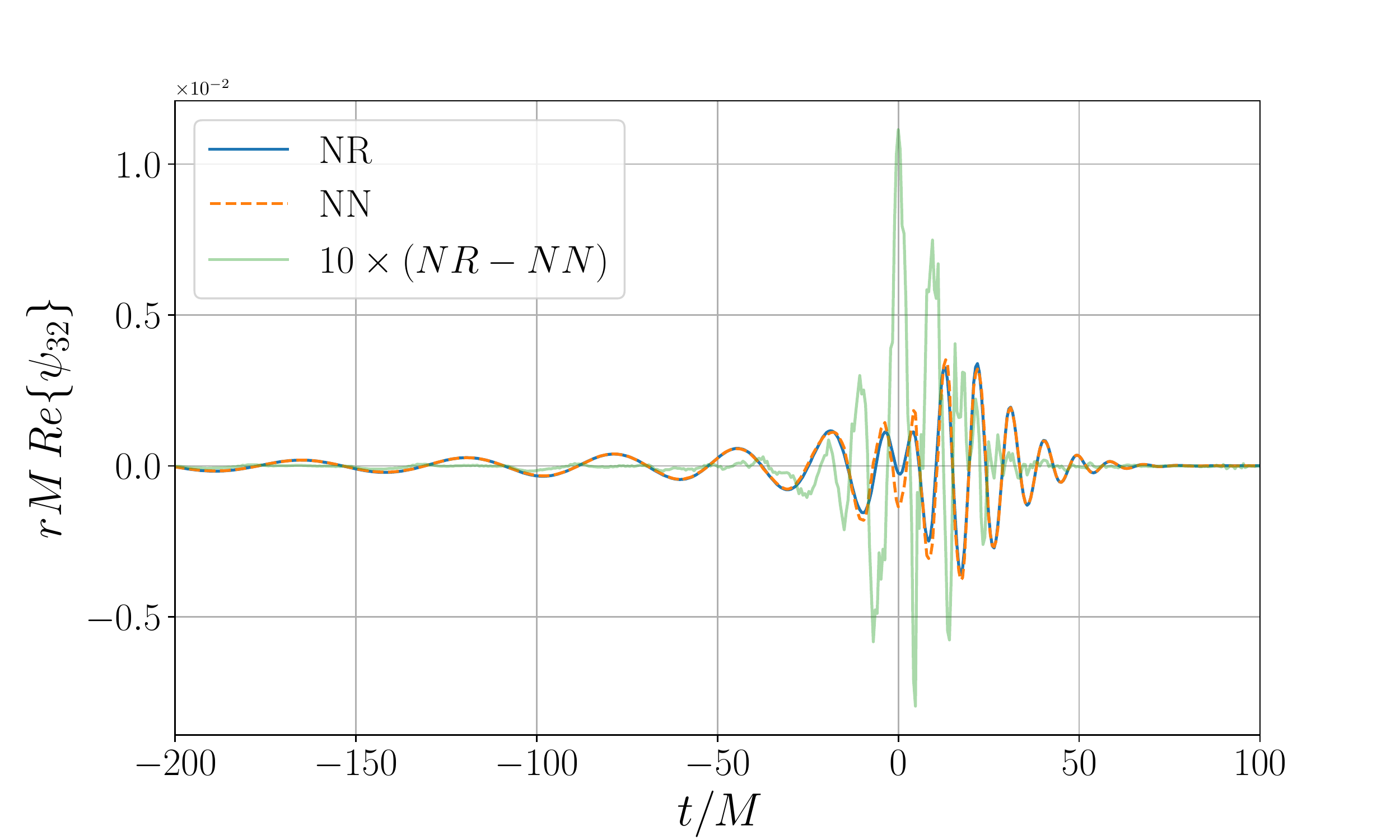}
    \caption{NR vs NN comparison of the $\psi_{32}$ from the simulation ID = 1989. $Mismatch = 2.7\times 10^{-3}$; $q = 4.0$, $\vec\chi_1 = (-0.50, -0.29, -0.55)$, $\vec\chi_2 = (0.02, -0.10, -0.79)$, and $e = 6.1\times 10^{-4}$.}
 \label{IM_plot}
\end{figure}

Still, some $\psi_{lm}$ modes revealed smaller amplitude magnitudes than the predictions, in the initial region of the merger, as seen in Fig. \ref{IM_plot}. These \emph{initial-merger} (IM) anomalies are present in $\psi_{lm}$ with $\ID_{(2,2)} = [1982,\, 1989]$, $\ID_{(3,2)} = [1982,\, 1989]$, $\ID_{[(3,3),\, (4,3),\, (4,4)]} = 1995$.

\begin{figure}[h]\centering
\hspace*{-0.2cm}
 \includegraphics[width=8.5cm]{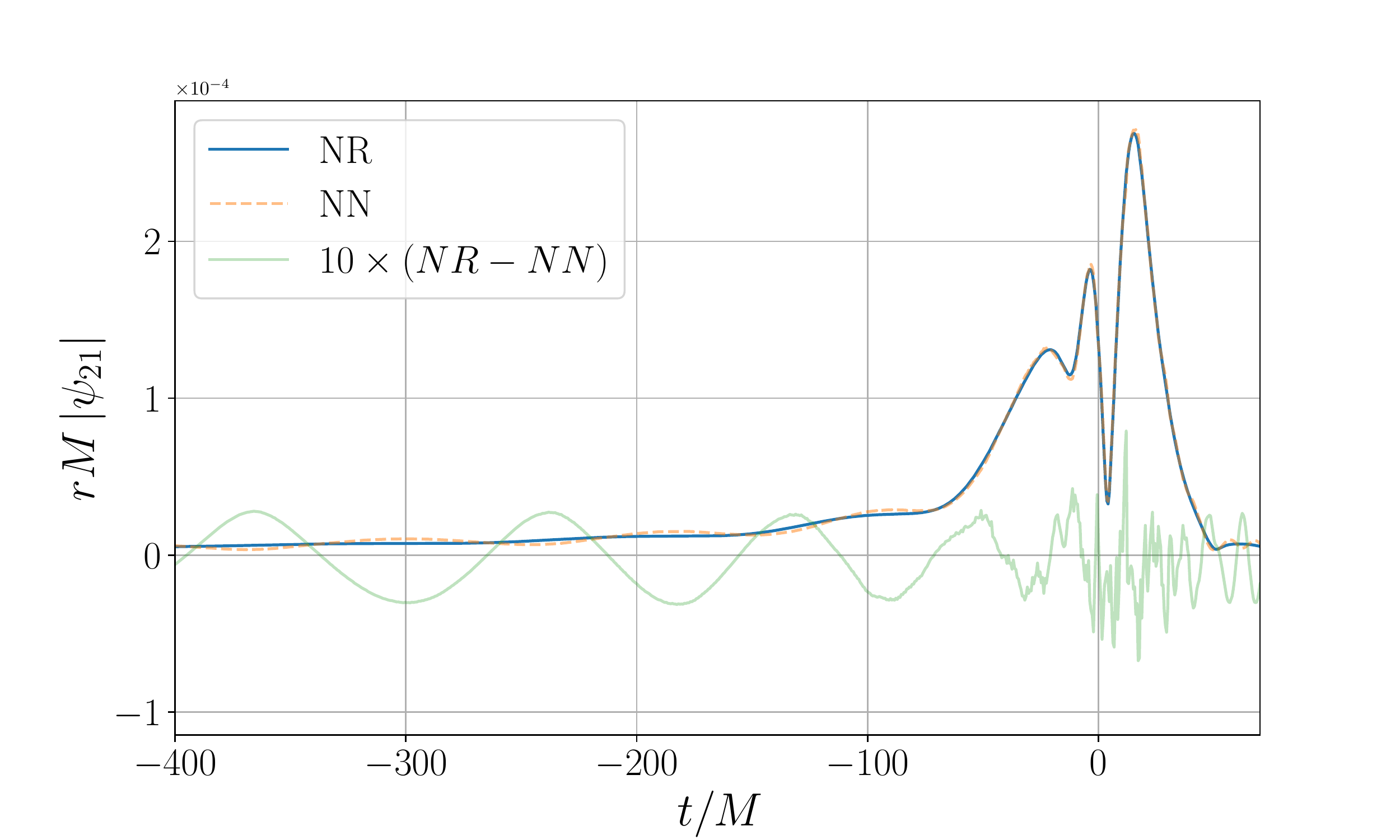}
 \caption{NR vs NN comparison of the $\psi_{21}$ from the simulation ID = 0292, showing a DI anomaly. $Mismatch = 3.87\times 10^{-3}$; $q = 3.0$, $\vec\chi_1 = (0.0, 0.0, 0.73)$, $\vec\chi_2 = (0.0, 0.0, -0.85)$, and $e = 1.5\times 10^{-4}$.}
 \label{DI_plot}
\end{figure}

In the $\psi_{lm}$ mode $(2, 1)$, we did not find MP or IM anomalies but
  discrepancies between NR and NN waveforms in phase and amplitude of the
  inspiral region in $\ID_{(2,1)} = [0013,\, 0252,\, 0292,\, 0388,\, 0513,\, 0464,\, 1508,\, 2126]$.
  The dephased-inspiral (DI) anomaly, as seen in Fig.~\ref{DI_plot},
  occurs in non-precessing binaries, in which we do not expect such oscillations in the waveform inspiral amplitudes.
  In this figure, we decreased the transparency of the NN prediction for better
  visualization of the oscillations between $[-400:-80] t/M$ in the NR waveform
  amplitude. This particular set of simulations was found with eccentricity
  $e > 10^{-4}$.

\begin{figure}[h]\centering
  \hspace*{-0.4cm}
  \includegraphics[width=8cm]{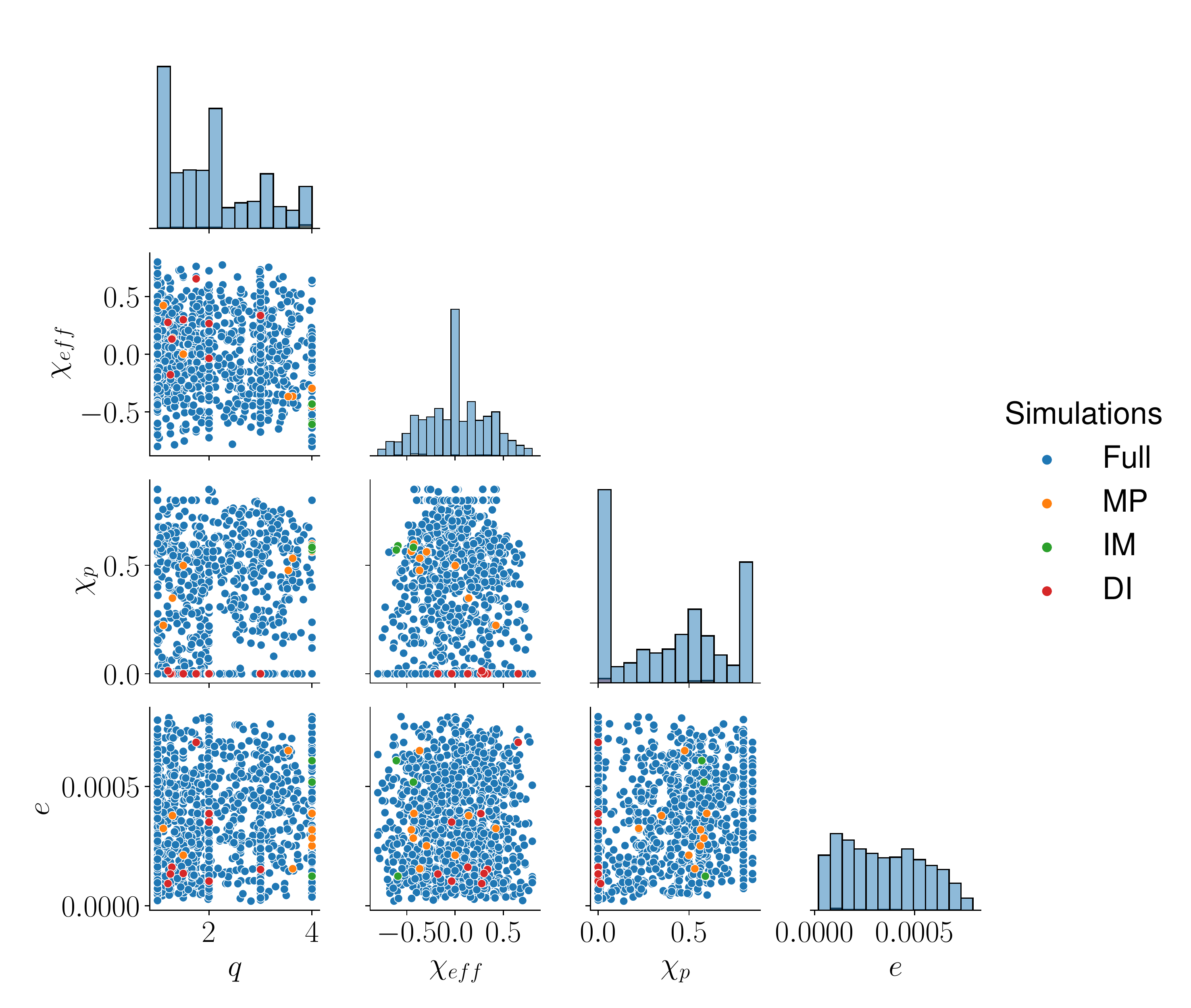}
  \caption{Analog to Fig.~\ref{fig:paramsf} and \ref{fig:paramsc}, but
      highlighting only the anomalies of the $\psi_4$ set, see Tab.~\ref{tab:anom} for
      acronyms.}
  \label{fig:paramspsi}
\end{figure}

The results of the anomaly search presented in this subsection are
summarized in Fig.~\ref{fig:paramspsi}, where IM anomaly appears only for
$q=4$, \emph{i.e.}, at the boundary of our waveform sample,
and DI anomaly is present only in non-precessing binaries which makes stronger the case for an actual anomaly.

\subsection{Anomaly detection statistics}
\label{ssec:statistics}

The anomalous waveforms found are mostly local morphological defects that differ from the waveform reproduced by WALDO as in Fig. \ref{IM_plot}. Still, we want to know whether our model has learned the waveform features corresponding to the entire parameter space of Fig. \ref{fig:paramsf} or it is creating such anomalies. For this, we simulate fake waveforms and check whether WALDO evaluates them as outliers.

To generate a fake waveform, we inject a Gaussian noise $n(t)$ of size $N_t$ into a randomly chosen $h_{lm}$ waveform from the dataset. So, we calculate the signal-to-noise ratio (SNR) from the signal $s_{lm} = h_{lm} + n$ as follows
\bq
    SNR_n = \frac{\left<h_{lm}|s_{lm}\right>}{\left<h_{lm}|h_{lm}\right>^{1/2}} \, ,
\eq
this time estimating the spectral noise density $S_n$ of $n$ and using it
  in the definition (Eq. \ref{eq:snr}) for the scalar product. All ``fake'' waveforms will receive the SNR value, in large,
  together with their ID number so that we can identify them after the mismatch evaluation.

\begin{figure}[h]\centering
  \hspace*{-0.4cm}
  \includegraphics[width=8.5cm]{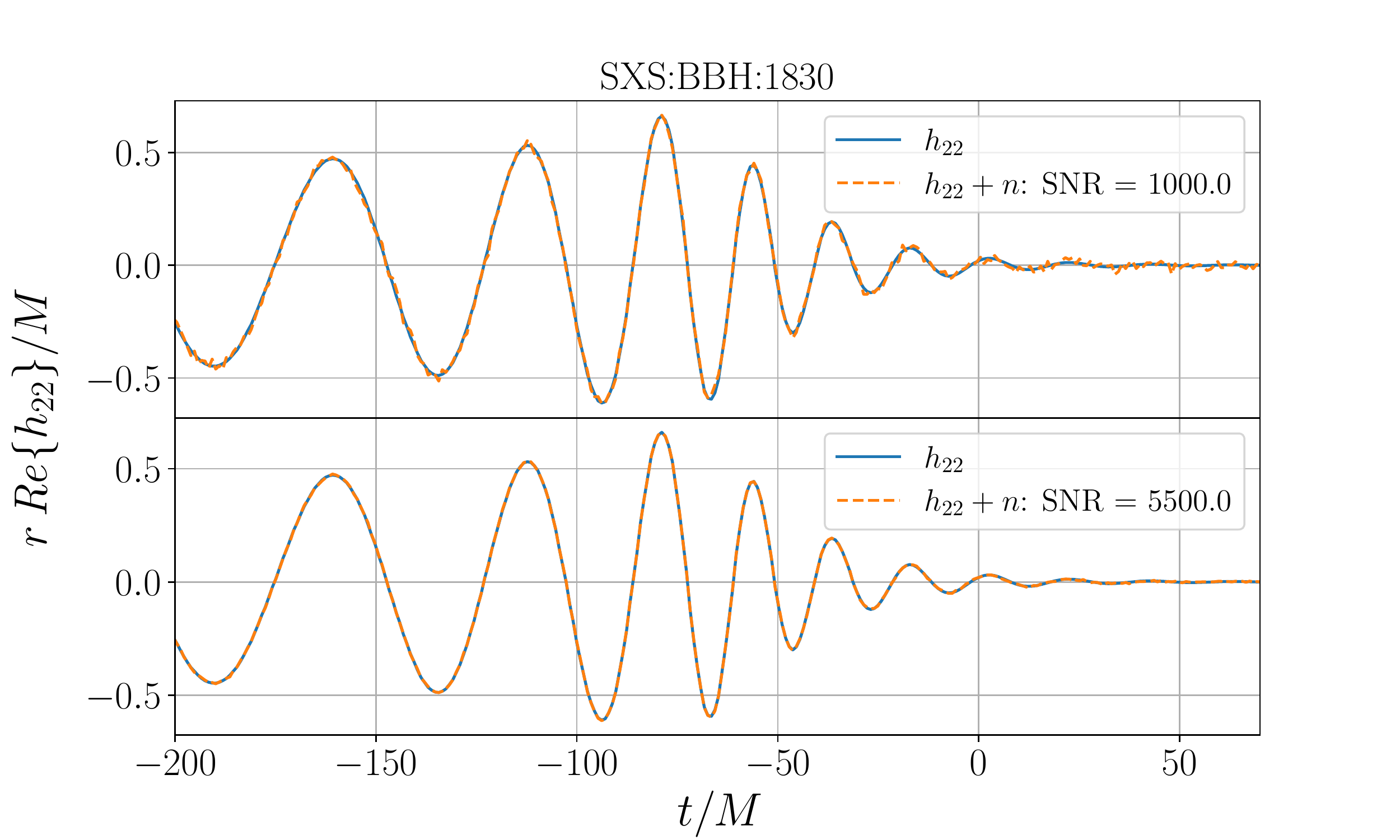}
  \caption{Waveform $h_{22}$ (blue line) from SXS:BBH:1830 simualtion. The plot above shows the noise injection of $SNR_n$ = 1000.0; below, an injection of $SNR_n$ = 5500.0 (dashed orange lines).}
  \label{fig:wfs_snr}
\end{figure}

We adopt an $SNR_n$ range $[1.0:10.0]\times 10^3$ for the experiment. Figure \ref{fig:wfs_snr} shows an example of a waveform with noise (dashed orange line) and without noise (blue line), for $SNR_n$ = 1000.0 on the plot above and $SNR_n$ = 5500.0 on the plot below.

To guarantee a dataset free from anomalous waveforms, we kept 30\% of the best waveforms of each mode $(l,\,m)$, which formed the \emph{clean-dataset} with 2904 waveforms.
We inject noise into 10 waveforms from the clean-dataset at random. 
With WALDO trained on the original dataset, we evaluate the clean-dataset and separate 5\% of the highest mismatch waveforms (18 outliers). 
Next, we compute how many waveforms from the 10 fake ones appeared among the 18 outlier waveforms.

\begin{figure}[h]\centering
  \hspace*{-0.4cm}
  \includegraphics[width=7.5cm]{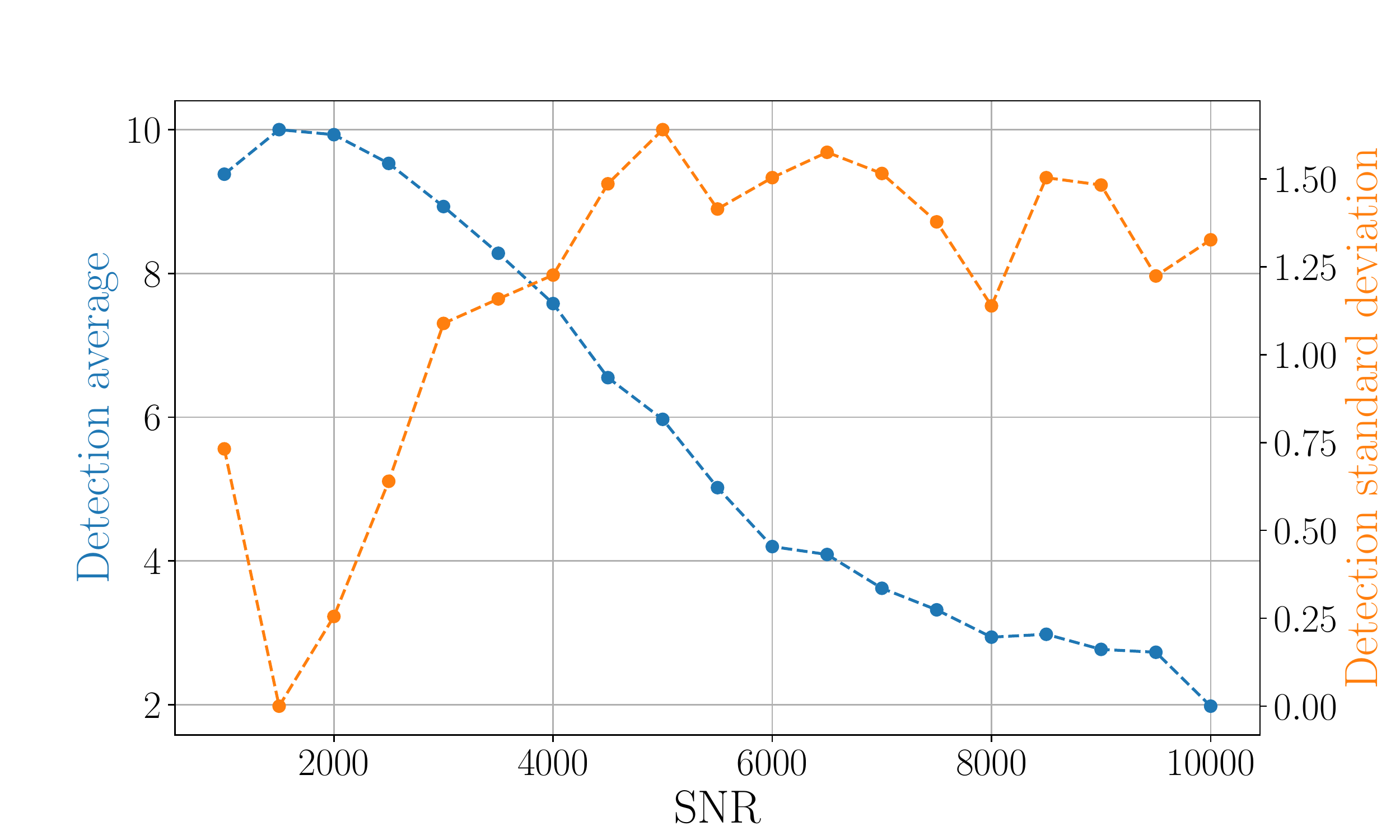}
  \caption{Detection average (blue line) and standard deviation (orange line) of fake waveforms per $SNR_n$ value.}
  \label{fig:detec}
\end{figure}

This experiment was repeated 100 times so that we have, in Fig. \ref{fig:detec}, the mean amount (left-hand side y-axis and blue line) and standard deviation (right-hand side y-axis and orange line) of fake waveforms found per each $SNR_n$ value. This result indicates that WALDO starts to present a doubtful detection (average of false waveforms found equal to 5 out of 10) for $SNR_n$ $\geq 5500.0$ (see fake waveform in the plot below of Fig. \ref{fig:detec}).

\section{Discussion}
\label{sec:concl}

\begin{table*}[htb]
\begin{tabular}{ll}
\hline
\textbf{Anomaly} & \textbf{Description} \\
\hline
\hline
(AR) Asymptotic-Ringdown &  Non-null asymptotic behavior during ringdown. \\
\hline
(LR) Lazy-Ringdown &  Ringdown late decay. \\
\hline
(RR) Rippled-Ringdown & Ripples in the dominant mode ringdown amplitude. \\
\hline
(SR) Short-Ringdown &  Ringdown with length below $70 M$. \\
\hline
(MP) Merger-Peak &  Higher amplitude around the merger peak.\\
\hline
(IM) Initial-Merger &  Shorter amplitude in the merger beginning.\\
\hline
(DI) Dephased-Inspiral & Oscillatory dephasing during inspiral. \\
\hline
\end{tabular}
\caption{Seven waveform anomalies detected by WALDO.}
\label{tab:anom}
\end{table*}

To assess the quality of Numerical Relativity data, and to identify problematic waveform candidates, we developed the Waveform
AnomaLy DetectOr (WALDO) wich allowed us to identify potentially anomalous waveforms both in the dominant and higher modes. We trained our model with 8046 waveforms 
-- considering the SXS center of mass (CoM) correction and the outermost extraction --
with a U-Net neural architecture and calculated the mismatch between the NR waveforms
and the NN predictions. By isolating the 1\% waveforms with highest mismatch, we
identified seven qualitatively different anomalies during the inspiral, merger, and ringdown stages. Table \ref{tab:anom} summarizes the anomaly categories.

We focus our anomaly search on the  $quantile > 0.99$ region of the mismatch,
and for higher modes with $(l \leq 4,\, l-1 \leq m \leq l)$.

The different anomalies found by WALDO are \emph{potential} inaccuracies in the waveform extraction or numerical evolution of field equations. We are not pointing out these simulations as definitively incorrect, but we are discussing a way to assess the quality of these data to identify
a small subset of the entire waveform catalog which may deserve a closer scrutiny.

While lacking analytic approximation approaches to check the consistency of the NR waveforms, as they only cover the inspiral and ringdown stages separately, we believe that the complete evaluation of the coalescence phases would only be possible -- as far as we know -- by correlating NR waveforms from different catalogs, which seems impractical.

We highlighted seven types of anomalous NN reconstructions whose patterns repeated for different simulations. From those, four types are undoubtedly morphological anomalies from simulation issues:
  \begin{enumerate}
    \item Asymptotic-Ringdown -- all ringdown waveform amplitudes from a compact binary coalescence source go asymptotically to zero around 100 solar mass after the amplitude peak;
    \item Short-Ringdown -- they are anomalous simulations concerning the catalog since they finish too early to complete an asymptotically zero amplitude;
    \item Rippled-Ringdown -- the ringdown amplitude of the dominant mode is analytically expected to decay smoothly, without mode mixing;
    \item Dephased-Inspiral -- found in $\psi_{21}$ waveforms, seems to be a mode mixing effect due to the binary CoM drift caused by the residual momentum of the simulation's initial condition \cite{Reisswig2011}.
  We stress that oscillations qualitatively similar to the ones in Fig. \ref{DI_plot}, while should not appear in non-precessing waveforms, are common in precessing ones, which are the majority in the catalog. Hence it is likely that NN learned from precessing waveforms to reproduce amplitude oscilations and output it also in a non-precessing case.
  \end{enumerate}
  The three remaining anomalies (Merger-Peak, Initial-Merger, and Lazy-Ringdown) are very local patterns in the merger-ringdown phases. IM and LR are the only two cases with simulations in the boundary of some parameter range: IM for equal masses and spin-aligned BBHs; and LR for mass-ratio $q=4$, but varied spin values.
  Since we found four different types of anomalies using our methodology, we could not fail to show these other three discrepancies between NR and NN waveforms.

We calculated the average of fake waveform detections created by noise injection (Sec. \ref{ssec:statistics}) for a statistical analysis of the anomalous waveform detection. This result indicates that the limit of the detection performance is for $SNR_n$ $\geq 5500.0$, which are fake waveforms with negligible noise (see Fig. \ref{fig:wfs_snr}).

The use of machine learning can be a great ally to extracting catalog features and reproducing waveforms with confidence. The present work intends to be a starting point for a more thorough investigation of NN applied to numerical waveforms.

As most of the anomalies appeared during the merger-ringdown stages, this may suggest the need for improvement of the \emph{adaptive mesh refinement} method of numerical simulations \cite{Szilgyi:2014}. Since the more refined the calculations during the collision of black holes, the more accurate the waveform during the merger.

We stress that for our analysis the dataset need to be as homogeneous
as possible in terms of astrophysical parameter space,
to avoid large mismatch values when dealing with anomaly-free waveforms because of
poor modeling. It is essential to remove simulations that have anomalies from the dataset
and re-train the NN to ensure that low-quality simulations do not
polllute the training set. Such anomalies can impair waveform modeling \cite{Khan2019,Taracchini2014,Blackman2015,Blackman2017} and interfere with analysis, such as the ringdown quasi-normal modes \cite{Leaver1985,Maggiore2008,Yang2012}.

We propose that WALDO can be applied to any time series, such as gravitational waves from binary neutron star and back hole-neutron star binary. Also, we suggest
to evaluate the quality of new simulations for the next generations of NR codes by comparing them with waveforms from well-established catalogs in the literature.

\begin{acknowledgments}
The authors thank the International Institute of Physics for hospitality
and support during most of this work.
We thank Michael Boyle, from the SXS team, for the valuable discussions.  
TP is supported by the Coordenação de Aperfei\c{c}oamento de Pessoal de
N\'\i vel Superior (CAPES) -- Graduate Research Fellowship. 
The work of RS is partly supported by CNPq under grant
310165/2021-0 and RS would like to thank ICTP-SAIFR FAPESP Grant
No. 2016/01343-7.
The authors thank the High Performance Computing Center (NPAD) at UFRN for
providing the computational resources necessary for this work.
\end{acknowledgments}


\bibliography{paper}

\end{document}